\documentclass{SolarPhysics}    
\usepackage[optionalrh]{spr-sola-addons}
\usepackage{epsfig}
\usepackage{url}
\usepackage{fixltx2e}
\usepackage[usenames]{color}

\newcommand{\ang}{\AA\ }
\newcommand{\gapprox}{\lower.4ex\hbox{$\;\buildrel >\over{\scriptstyle\sim}\;$}}
\newcommand{\lapprox}{\lower.4ex\hbox{$\;\buildrel <\over{\scriptstyle\sim}\;$}}
\newcommand{\arcsec}{\hbox{$^{\prime\prime}$}}

\def\ang{\AA}

\def\aap  {{\sl Astron. Astrophys.}\ }   
\def\apj  {{\sl Astrophys. J.}\ }        
\def\aj   {{\sl Astronom. J.}\ } 	 
\def\jastp{{\sl J. Atmos. Solar-Terr. Phys.}\ } 
\def\sp   {{\sl Solar Phys.}\ }          
\def\ssr  {{\sl Space Science Rev.}\ }   

\begin{document}
\begin{article}
\begin{opening}
\title{Solar Stereoscopy with STEREO/EUVI A and B spacecraft
	from small ($6^\circ$) to large $(170^\circ)$ spacecraft 
	separation angles}

\author{Markus J. Aschwanden \sep Jean-Pierre W\"ulser \sep 
  	Nariaki Nitta \sep James Lemen}
\runningauthor{M.J. Aschwanden \textit{et al.}}
\runningtitle{Solar Stereoscopy}

\institute{Solar and Astrophysics Laboratory,
	Lockheed Martin Advanced Technology Center, 
        Dept. ADBS, Bldg.252, 3251 Hanover St., Palo Alto, CA 94304, USA; 
        (e-mail: \url{aschwanden@lmsal.com})}

\date{Received 22 Nov 2011; Revised 19 March 2012; Accepted ...}

\begin{abstract}
We performed for the first time stereoscopic triangulation of coronal
loops in active regions over the entire range of spacecraft separation
angles ($\alpha_{sep}\approx 6^\circ, 43^\circ, 89^\circ, 127^\circ$,
and $170^\circ$). The accuracy of stereoscopic correlation depends mostly 
on the viewing angle with respect to the solar surface for each spacecraft,
which affects the stereoscopic correspondence identification of loops in
image pairs. From a simple theoretical model we predict an optimum range
of $\alpha_{sep} \approx 22^\circ-125^\circ$, which is also experimentally
confirmed. The best accuracy is generally obtained when an active region
passes the central meridian (viewed from Earth), which yields a symmetric
view for both STEREO spacecraft and causes minimum horizontal foreshortening. 
For the extended angular range of $\alpha_{sep}\approx 6^\circ-127^{\circ}$ we
find a mean 3D misalignment angle of $\mu_{PF} \approx 21^\circ-39^\circ$
of stereoscopically triangulated loops with magnetic potential field models,
and $\mu_{FFF} \approx 15^\circ-21^\circ$ for a force-free field model,
which is partly caused by stereoscopic uncertainties $\mu_{SE} \approx
9^\circ$. We predict optimum conditions for solar stereoscopy during the time 
intervals of 2012--2014, 2016--2017, and 2021--2023.
\end{abstract}


\end{opening}

\section{		Introduction		}

Ferdinand Magellan's expedition was the first that completed the 
circumnavigation of our globe during 1519-1522, after discovering the
{\sl Strait of Magellan} between the Atlantic and Pacific ocean in
search for a westward route to the ``Spice Islands'' (Indonesia),
and thus gave us a first $360^\circ$ view of our planet Earth. 
Five centuries later, NASA has sent two spacecraft of the STEREO
mission on circumsolar orbits, which reached in 2011 
vantage points on opposite sides of the Sun that give us a first
$360^\circ$ view of our central star. Both discovery missions are of
similar importance for geographic and heliographic charting, and the
scientific results of both missions rely on geometric triangulation. 

The twin STEREO/A(head) and B(ehind) spacecraft (Kaiser et al.~2008), 
launched on 2006 October 26, 
started to separate at end of January 2007 by a lunar swingby and became 
injected into a heliocentric orbit, one propagating ``ahead'' and the 
other ``behind'' the Earth, increasing the spacecraft separation angle 
(measured from Sun center) progressively by about $45^\circ$ per year.
The two spacecraft reached the largest separation angle of $180^\circ$ 
on 2011 February 6. A STEREO SECCHI COR1-A/B intercalibration was
executed at $180^\circ$ separation (Thompson et al.~2011).
Thus, we are now in the possession of imaging data from 
the two STEREO/EUVI instruments (Howard et al.~2008; W\"ulser et al.~2004) 
that cover the whole range from smallest 
to largest stereoscopic angles and can evaluate the entire angular range
over which stereoscopic triangulation is feasible. It was anticipated
that small angles in the order of $\approx 10^\circ$ should be most
favorable, similar to the stereoscopic depth perception by eye, while
large stereoscopic angles that are provided in the later phase of the
mission would be more suitable for tomographic 3D reconstruction. 

The first stereoscopic triangulations using the STEREO spacecraft have
been performed for coronal loops in active regions, observed on 2007 May 9 
with a separation angle of $\alpha_{sep}=7.3^\circ$ (Aschwanden et al.~2008) 
and observed on 2007 June 8 with $\alpha_{sep}=12^\circ$ (Feng et al.~2007). 
Further stereoscopic triangulations have been applied to oscillating loops 
observed on 2007 June 26 with a stereoscopic angle of $\alpha_{sep}=15^\circ$
(Aschwanden 2009), to polar plumes observed on 2007 Apr 7 with 
$\alpha_{sep}=3.6^\circ$ (Feng et al.~2009), to an erupting filament
observed on 2007 May 19 with $\alpha_{sep}=8.5^\circ$ (Liewer et al.~2009), 
to an erupting prominence observed on 2007 May 9 with 
$\alpha_{sep}=7.3^{\circ}$ (Bemporad 2009), and to a rotating,
erupting, quiescent polar crown prominence observed on
2007 June 5-6 with $\alpha_{sep}=11.4^\circ$ (Thompson 2011). 
Thus, all published stereoscopic triangulations
have been performed within a typical (small) stereoscopic angular range
of $\alpha_{sep} \approx 3^\circ-15^\circ$, as it was available during
the initial first months of the STEREO mission. 
The largest stereoscopic angle used for triangualtion of coronal loops
was used for active region 10978, observed on 2007 December 11, with
a spacecraft separation of $\alpha_{sep}=42.7^\circ$
(Aschwanden and Sandman 2010; Sandman and Aschwanden 2011),
which produced results with similar accuracy as those obtained from 
smaller stereoscopic angles. So there exists also an intermediate
rangle of aspect angles that can be used for stereoscopic triangulation.

However, nothing is known whether stereoscopy is also feasible at
large angles, say in the range of $\alpha_{sep} \approx 50^{\circ}-180^\circ$,
and how the accuracy of 3D reconstruction depends on the aspect angle,
in which range the stereoscopic correspondence problem is intractable,
and whether stereoscopy at a maximum angle near $\alpha_{sep} 
\lapprox 180^{\circ}$ is equally feasible as for $\alpha_{sep} 
\gapprox 0^\circ$ for optically thin structures (as it is the case
in soft X-ray and EUV wavelengths), due to the $180^\circ$ symmetry 
of line-of-sight intersections. In this study we are going to explore
stereoscopic triangulation of coronal loops in the entire range of
$\alpha_{sep} \approx 6^\circ - 170^\circ$ and quantify the accuracy
and quality of the results as a function of the aspect angle.

Observations and data analysis are reported in Section 2, while a
discussion of the results is given in Section 3, with conclusions
in Section 4. 

\begin{figure}
\centerline{\includegraphics[width=1.0\textwidth]{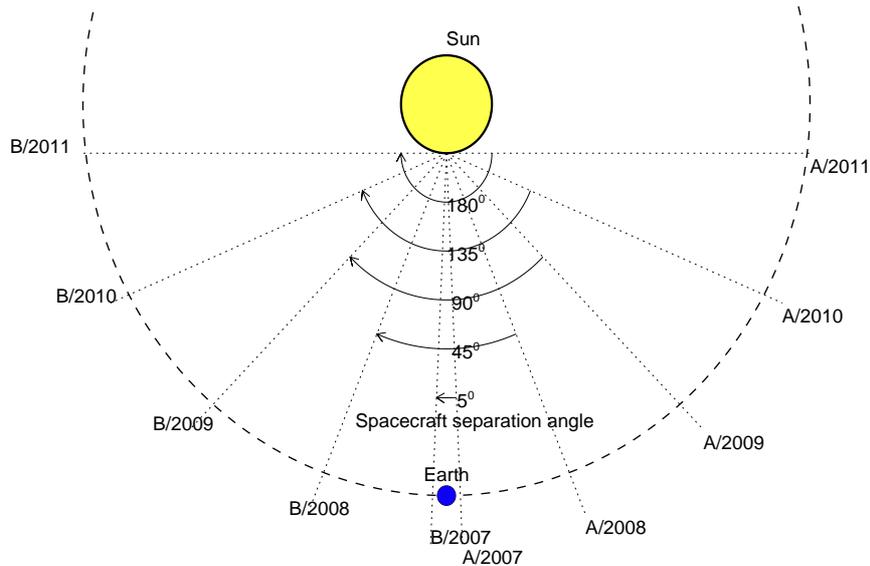}}
\caption{Schematic figure of the spacecraft orbits of STEREO/A and B
relative to Earth, with the spacecraft separation angles 
$\alpha_{sep}=\alpha_A-\alpha_B$ indicated approximately at the 
beginning of the years, ranging from $\approx 5^\circ$ in April 2007 to
$\approx 180^\circ$ in February 2011.}
\end{figure}

\section{	OBSERVATIONS AND DATA ANALYSIS			}

\subsection{		Observations				}

We select STEREO observations at spacecraft separation
angles with increments of $\approx 45^\circ$ over the range of
$\alpha_{sep} \approx 0^\circ$ to $\approx 180^\circ$, which corresponds
to time intervals of about a year during the past mission lifetime 2007--2011.
A geometric sketch of the spacecraft positions STEREO/A+B relative to
the Earth-Sun axis is shown in Fig.~1.
Additional constraints in the selection are: (i) The presence of a relatively
large prominent active region; (ii) a position in the field-of-view of 
both spacecraft (since the mutual coverage overlap drops progressively from 
$180^\circ$ initially to $0^\circ$ during the first 4 years of the mission); 
(iii) a time near the central meridian passage of an active region 
viewed from Earth (to minimize confusion by foreshortening); and (iii) the 
availability of both STEREO/EUVI/A+B and calibrated SOHO/MDI data.  
The selection of 5 datasets is listed in Table 1, which includes
the following active regions: (1) NOAA 10953 observed on 2007 April 30 
(also described
in DeRosa et al.~2009; Sandman et al.~2009, Aschwanden and Sandman 2010;
Sandman and Aschwanden 2011, Aschwanden et al.~2012),
(2) NOAA region 10978 observed on 2007 December 11 (also described in
Aschwanden and Sandman 2010, Aschwanden et al.~2012, and subject
to an ongoing study by Alex Engell and Aad Van Ballegooijen,
private communication), (3) NOAA 11010 observed on 2009 Jan 12,
(4) NOAA 11032 observed on 2009 Nov 21, and 
(5) NOAA 11127 observed on 2010 Nov 23. 
This selection covers spacecraft separation angles of
$\alpha_{sep} \approx 6^\circ, 43^\circ, 89^\circ, 127^\circ$, and
$170^\circ$. 

For each of the 5 datasets we stacked the images during a time interval
of 20--30 minutes in order to increase the signal-to-noise ratio of
the EUVI images. During the first 3 years (2006-2008) the nominal cadence 
of 171 \ang\ images was 150 s, which yields 8 stacked images per 
20 minute interval. Later in the mission, the highest cadence was
chosen for the 195 \ang\ wavelength, but dropped from 150 s to 300 s 
due to the reduced telemetry rate at larger spacecraft distances, which
yields 6--12 stacked images per 30 minute interval. In one case
(2010 Nov 23) the cadence in EUVI/A and B are not equal, either
due to data loss or different telemetry priorities (see time intervals
and number of stacked images in Table 1). The solar rotation during the
time interval of stacked image sequences was removed to first order 
by shifting the images by an amount corresponding to the rotation rate 
at the extracted subimage centers. 

\begin{table}
\caption{Data selection of 5 active regions observed with STEREO/EUVI
and SOHO/MDI.}
\begin{tabular}{llllrr}
\hline
Active & Observing & Observing & Wave-  & Number of      & Spacecraft \\ 
Region & date      & times     & length & stacked images & separation \\
       &           & (UT)      & (A)    & $N_A$, $N_B$   & angle (deg)\\
\hline
10953 (E23S10)  &2007-Apr-30    &23:00-23:20  &171   &  8,  8 &   6.1$^\circ$\\
10978 (E14S01)  &2007-Dec-11    &16:30-16:50  &171   &  8,  8 &  42.7$^\circ$\\
11010 (E05N18)  &2009-Jan-12    &00:30-01:00  &171   &  6,  6 &  89.3$^\circ$\\
11032 (W03N16)  &2009-Nov-21    &00:30-01:00  &195   & 12, 12 & 126.9$^\circ$\\ 
11127 (W08N25)  &2010-Nov-23    &00:30-01:00  &195   & 12,  6 & 169.4$^\circ$\\ 
\hline
\end{tabular}
\end{table}

\subsection{	Stereoscopic Triangulation		}

The geometric principles of stereoscopic triangulation are described in
Aschwanden et al.~(2008; Sections 3.1 and 3.2 therein) for the general 
case of different spacecraft distances $d_A$ and $d_B$ from the Sun. 
The given formulas work correctly up to spacecraft separation angles of
$\alpha_{sep} \lapprox 90^\circ$, but there is a sign ambiguity for
larger separation angles. However, using the publicly available SSW/IDL 
software in the framework of the {\sl Wold Coordinate System (WCS)}
(Thompson 2006), we can transform a pair of STEREO/A+B images into
epipolar coordinates (Inhester 2006), by coaligning to the same Sun 
center position, derotating the spacecraft roll angles, and 
rescaling to the same solar distance $d_A$), where stereoscopic 
triangulation is most straightforward. If $(x_A, y_A)$ are the cartesian 
coordinates of a loop position in a STEREO/A image (in units of solar
radii measured from Sun center), the relationship between the heliocentric 
longitudes $(l_A, l_B)$ and latitudes $(b_A, b_B)$ in the images $A$ and
$B$, and the cartesian coordinates $(x_B, y_B)$ in the image $B$ are
(in the far-field approximation),
\begin{equation}
	\begin{array}{ll}
		b_A &= \arcsin{(y_A/r)}		\\
		l_A &= \arcsin{[x_A/(r \ \cos{b_A})]}\\
		l_B &= l_A + \alpha_{sep}	\\
		b_B &= b_A			\\
		x_B &= r \ \sin(l_B) \cos(b_B)	\\
		y_B &= y_A			
	\end{array} \ .
\end{equation}
where $r=1+h$ is the stereoscopically triangulated distance from Sun center 
(in units of solar radii). 
The conversion of image $A$ pixel coordinates $(i_A, j_A)$ into dimensionless 
Sun center coordinates $(x_A, y_A)$ is, with $r_{pix}$ the solar radius in
units of EUVI pixels sizes,
\begin{equation}
	\begin{array}{ll}
		x_A &= (i_A - i_0)/r_{pix} \\
		y_A &= (j_A - j_0)/r_{pix} 
	\end{array} \ ,
\end{equation}
and the back-transformation into image $B$ pixel coordinates is,
\begin{equation}
	\begin{array}{ll}
		i_B &= i_0 + x_B \ r_{pix} \\
		j_B &= j_0 + y_B \ r_{pix} 
	\end{array} \ ,
\end{equation}
where $(i_0, j_0)$ are the pixel coordinates of the Sun center
in the co-registered epipolar STEREO image pair.
There is no sign ambiguity in the coordinate transformation,
as long as the triangulated feature is in front of the visible
hemisphere (and plane-of-sky through Sun center) for each spacecraft, 
or more specifically, if the triangulated positions have positive
values $z_A > 0$ and $z_B > 0$ in each spacecraft coordinate system.

\subsection{	Magnetic Field Models 			}

Since the two STEREO/A+B spacecraft have an almost symmetric separation
angle in east and west direction with respect to Earth, their overlapping
field-of-view is always centered closely to the Sun's central meridian
for spacecraft separation angles of $\alpha_{sep} \le 180^\circ$, 
and thus we have always also a magnetogram from an Earth-bound 
satellite available, such as from SOHO/MDI, during the considered
time period of 2006-2011. We make use of the {\sl Michelson Doppler
Imager (MDI)} daily magnetic field synoptic full-disk data, which
are taken every 96 minutes, and thus are near-simultaneous with the
STEREO/EUVI images within $\approx 1$ hour. The maximum magnetic
field strengths of the 5 analyzed active regions are listed in Table 2, 
reaching up $B \approx 3100$ G. 

We calculate a magnetic potential-field model of each analyzed active
region by the method of buried unipolar magnetic charges, which is
described to first approximation in Aschwanden and Sandman (2010),
and with a higher accuracy including the curvature of the solar
surface in Aschwanden et al.~(2012). Essentially, a line-of-sight
magnetogram $B_z(x,y)$ is decomposed into a number (typically
$n_c=200$) components of buried magnetic charges, each one parameterized
with 4 parameters ($B_j, x_j, y_i, z_j)$ that characterize the surface
field strength $B_j$ and the 3D position of the magnetic charge below 
the photospheric surface.
The coronal potential field ${\bf B}({\bf x})$ is calculated from the
superposition of all $N_m$ magnetic charges, 
\begin{equation}
	{\bf B}({\bf x}) = \sum_{j=1}^{N_m} {\bf B}_j({\bf x})
	= \sum_{j=1}^{N_m}  B_j
	\left({d_j \over r}\right)^2 {{\bf r} \over r} \ ,
	\end{equation}
with ${\bf r}=[(x-x_j), (y-y_j), (z-z_j)]$ being the distance of
a coronal position $[x, y, z]$ from the magnetic charge $j$, 
${\bf r}_j=(x_j,y_j,z_j)$ is the subphotospheric 
position of the buried charge, and $d_j=\sqrt{1-x_j^2-y_j^2-z_j^2}$
is the depth of the magnetic charge.

We calculate magnetic field lines by starting at photospheric footpoints
and extrapolating along the local magnetic field vector 
${\bf B}({\bf x})=(B_x, B_y, B_z)$ in steps of $\Delta s=0.004$ solar 
radii. We initiate the field line footpoints in a regular grid
(of say $50 \times 50$) footpoints positions, but plot only those 
field lines that have a photospheric field strength above some threshold, 
say $B \ge 100$ G. 

For the first three active regions where we have sufficient constraints
by stereoscopic loops ($N_{loops} \gapprox 20$), we calculate also a
{\sl nonlinear force-free field (NLFFF)} solution according to a new
code based on an analytical approximation of divergence-free and
force-free fields that includes an azimuthal magnetic field component 
with vertical twist and is accurate to second-order (of the force-free
parameter $\alpha$). This new NLFFF code uses the constraints of a 
line-for-sight magnetogram to define the potential field (Eq.~4) and
is suitable for fast forward-fitting to coronal field constraints, such
as the stereoscopically triangulated 3D loop coordinates as calculated
here. The analytical theory of this new NLFFF code is described in
Aschwanden (2012), the numerical code and tests in Aschwanden and
Malanushenko (2012), and first solar applications to four stereoscopically
observed active regions in Aschwanden et al.~(2012). 

\subsection{	Data Analysis and Results			}

In Figs.~2--6 we present the results of stereoscopically triangulated
loops for the 5 analyzed active regions and compare them with
magnetic potential field and force-free field models. 
In each of the Figures we use the
following layout of panels: Partial STEREO/A and B images are shown
in the top panels, each one with a field-of-view that encompasses
the active region of interest. The STEREO/A and B images are shown
for the brightness on a logarithmic color scale (top panels), as well as
in a highpass-filtered version (by unsharp masking, i.e., by subtracting
a $3\times 3$ boxcar-smoothed image from the original image) to enhance
the loop structures (panels in second row). The SOHO/MDI magnetogram
is shown in the bottom right panel, which has a different field-of-view 
from an Earth-bound vantage point. In addition we show a side view of
the active region by rotating the magnetogram view by $90^\circ$ to
the north (bottom left panel). In all panels, the projections of
stereoscopically triangulated loops are shown in blue color, and
the magnetic potential field lines in red color. 

\begin{table}
\caption{Data analysis of selection 5 active regions.}
\begin{tabular}{llllll}
\hline
Active & Number & Misalignment  & Misalignment  & Stereoscopy & Magnetic \\ 
Region & of     & angle         & angle         & error       & field strength\\
       & loops  & $\mu_{PF}$ (deg) & $\mu_{FFF}$ (deg)& $\mu_{SE}$ (deg)&B(Gauss)\\
\hline
10953  &100     &  27.7$^\circ$ & $19.8^\circ$ &  9.4$^\circ$  &[-3134,+1425]\\
10978  & 52     &  20.8$^\circ$ & $15.1^\circ$ &  8.9$^\circ$  &[-2270,+2037]\\
11007  & 20     &  36.2$^\circ$ & $16.5^\circ$ &  ...          &[ -737,+1342]\\ 
11035  & 15     &  29.8$^\circ$ & $17.4^\circ$ &  ...          &[ -997, +774]\\ 
11161  &  5     &  39.4$^\circ$ & $21.2^\circ$ &  ...          &[-2033, +884]\\ 
\hline
\end{tabular}
\end{table}

The first case is active region NOAA 10953 (Fig.~2), where we display
the same 100 loop segments that have been triangulated in an earlier
study (Aschwanden and Sandman 2010). The spacecraft separation angle
is $\alpha_{sep}=6.1^\circ$ and the almost identical direction of the
line-of-sights of both STEREO/A and B spacecraft makes it easy to
identify the corresponding loops in A and B, and thus the triangulation
is very reliable. Note that the height range where discernable loops
can be traced in the highpass-filtered images is about $h_{max}\approx
0.1$ solar radii (or $\approx 70$ Mm), which is commensurable with the
hydrostatic density scale height expected for a temperature of 
$T=1.0$ MK that corresponds to the peak sensitivity of the EUVI 171 
\ang\ filter. This is particularly well seen in the side view shown
in the bottom left panel in Fig.~2. A measurement of the mean 
misalignment angle averaged over 10 positions of the 100 reconstructed loops
with the local magnetic potential field shows a value of
$\mu_{PF}=27.7^\circ$ (Table 2), similar to earlier work
(Aschwanden and Sandman 2010; Sandman and Aschwanden 2011). 
However, forward-fitting of a nonlinear force-free field model
reduces the misalignment to $\mu_{FFF}=19.8^\circ$, which implies
that this active region is slightly nonpotential.
The remaining misalignment is attributed to at least two reasons, partially
to inadequate parameterization of the force-free field model, and partially to
stereoscopic measurement errors $\alpha_{SE}$ due to misidentified loop 
correspondences and limited spatial resolution. An empirical estimate
of the stereoscopic error was devised in Aschwanden and Sandman (2010),
based on the statistical non-parallelity of closely-spaced triangulated 
loop 3D trajectories, which yielded for this case a value of 
$\alpha_{SE}=9.4^\circ$. In summary, we find that this active region
is very suitable for stereoscopy, allows to discern a large number (100)
of loops, minimizes the stereoscopic correspondence problem
due to the small ($\alpha_{sep}=6.1^\circ$) spacecraft separation angle,
displays a moderate misalignment angle and stereoscopic 
measurement error ($\mu_{SE}=9.4^\circ$). This well-defined case will 
serve as a reference for stereoscopy at larger angles.

\begin{figure}
\centerline{\includegraphics[width=1.0\textwidth]{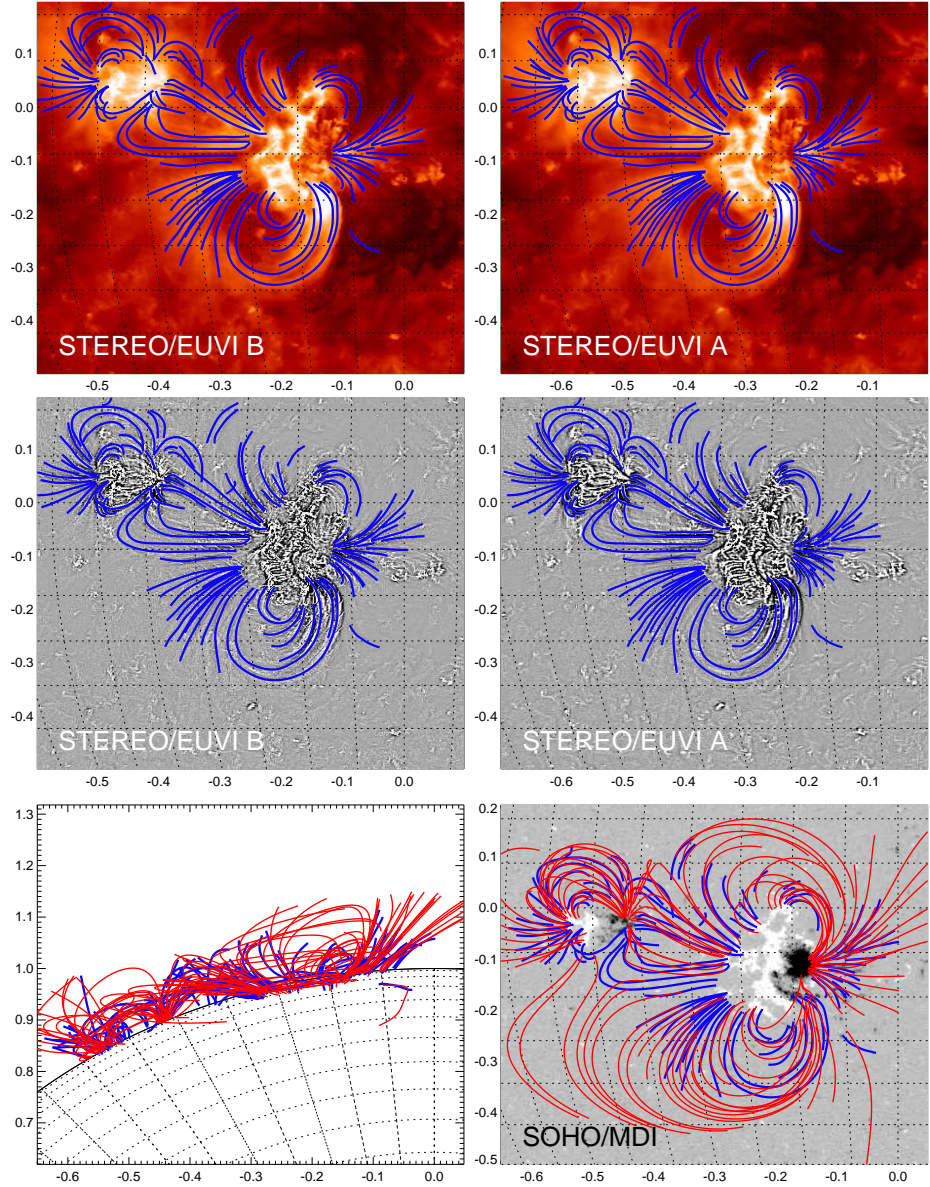}}
\caption{STEREO/EUVI spacecraft A (top right) and spacecraft B image 
(top left), of active region AR 10953, observed on 2007 Apr 30, 
23:00 UT in the 171 \ang\ wavelength, with a spacecraft separation 
angle of $\alpha_{sep}=6.1^\circ$. The images are highpass-filtered 
to enhance loop structures (middle left and right panels). 
A near-simultaneous SOHO/MDI magnetogram is shown (bottom right),
overlaid with the stereoscopically triangulated loops (blue curves) 
and magnetic field lines computed with a nonlinear force-free model
(red curves),
viewed from the direction of Earth or SOHO/MDI (bottom right), and 
rotated by $90^0$ to the north (bottom left).}
\end{figure}

\begin{figure}
\centerline{\includegraphics[width=1.0\textwidth]{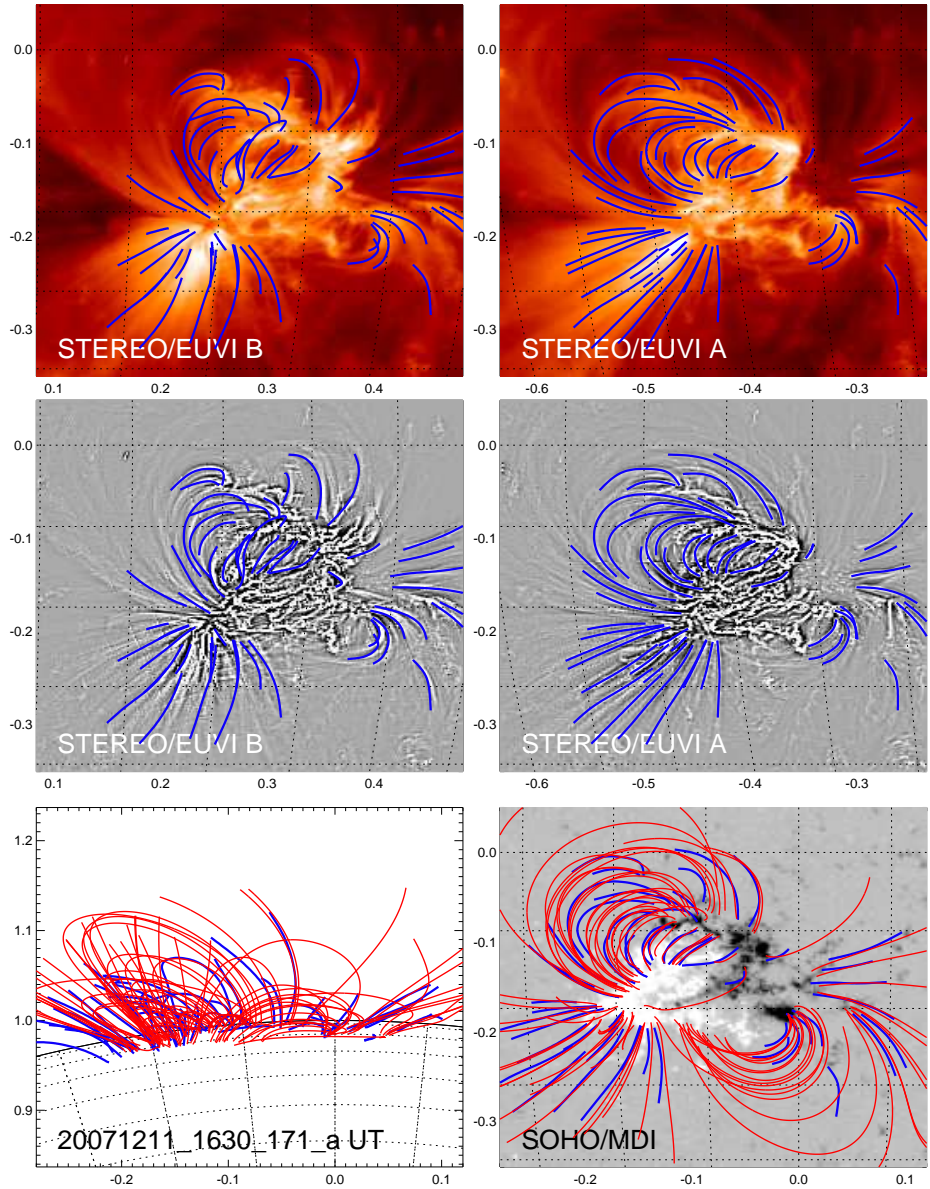}}
\caption{Active region AR 10978, observed on 2007 Dec 11, 16:30 UT 
in the 171 \ang\ wavelength, with a spacecraft separation angle of 
$\alpha_{sep}=42.7^\circ$.  
A SOHO/MDI magnetogram is shown (bottom right),
overlaid with the stereoscopically triangulated loops (blue curves) 
and magnetic field lines computed with a nonlinear force-free model
(red curves).}
\end{figure}

\begin{figure}
\centerline{\includegraphics[width=1.0\textwidth]{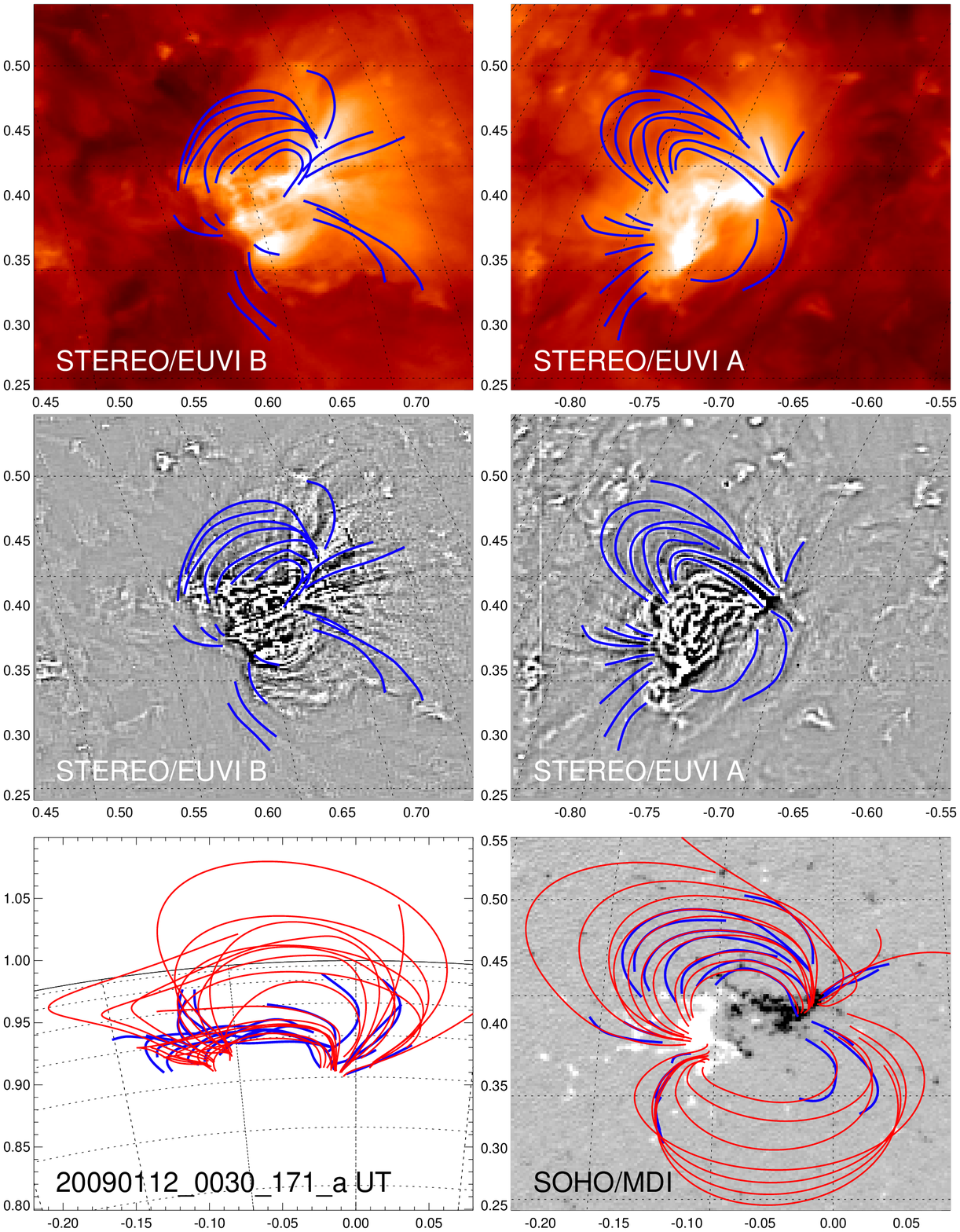}}
\caption{Active region AR 11010, observed on 2009 Jan 12, 00:30 UT 
in the 171 \ang\ wavelength, with a spacecraft separation angle of 
$\alpha_{sep}=89.3^\circ$.  
A SOHO/MDI magnetogram is shown (bottom right),
overlaid with the stereoscopically triangulated loops (blue curves) 
and magnetic field lines computed with a nonlinear force-free model
(red curves).}
\end{figure}

\begin{figure}
\centerline{\includegraphics[width=1.0\textwidth]{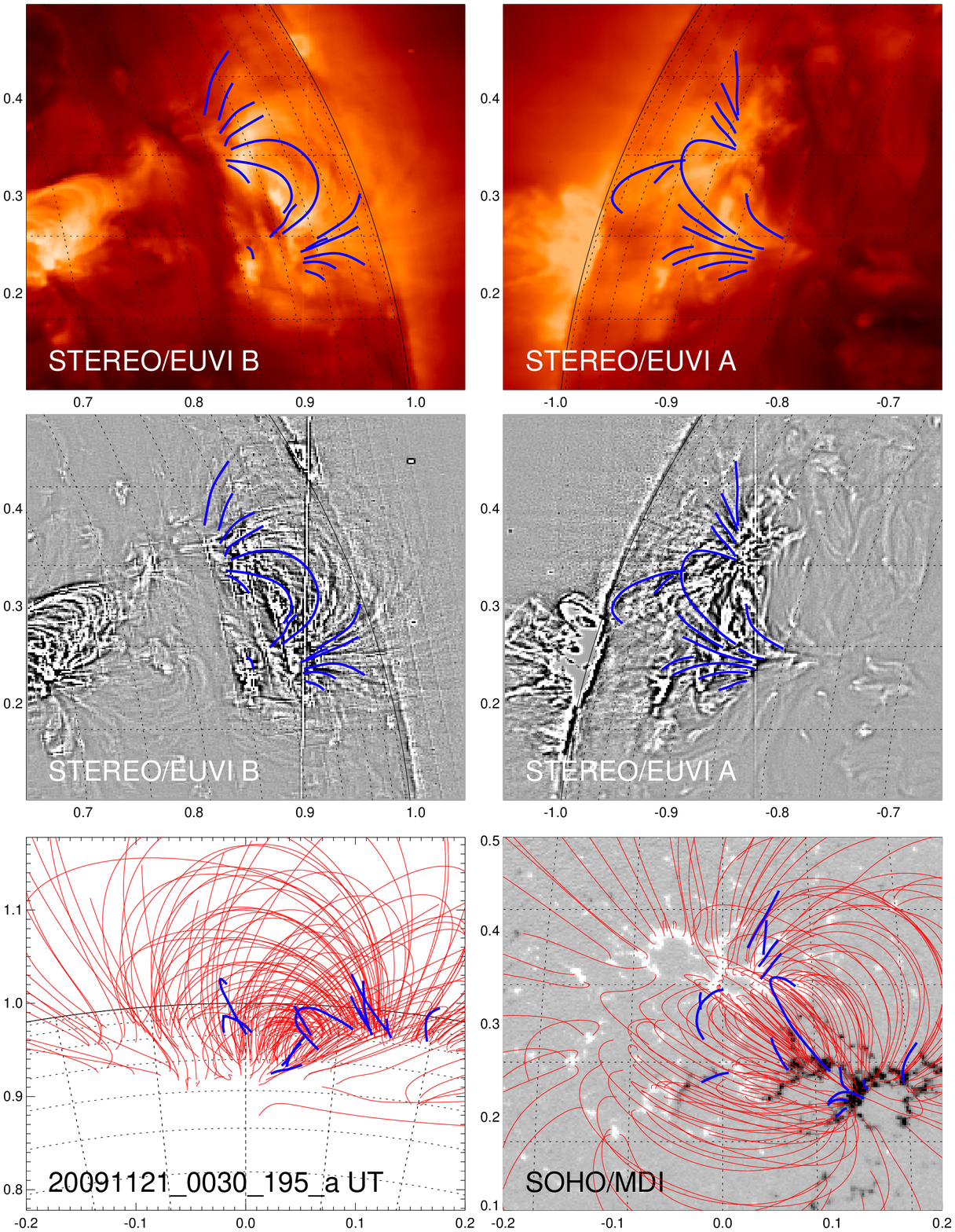}}
\caption{Active region AR 11032, observed on 2009 Nov 21, 00:30 UT 
in the 195 \ang\ wavelength, with a spacecraft separation angle of 
$\alpha_{sep}=126.9^\circ$. A SOHO/MDI magnetogram is shown, 
overlaid with the stereoscopically triangulated loops (blue curves) 
and magnetic field lines computed with a potential field model.
Magnetic field lines have a footpoint threshold of $B>100$ G.}
\end{figure}

\begin{figure}
\centerline{\includegraphics[width=1.0\textwidth]{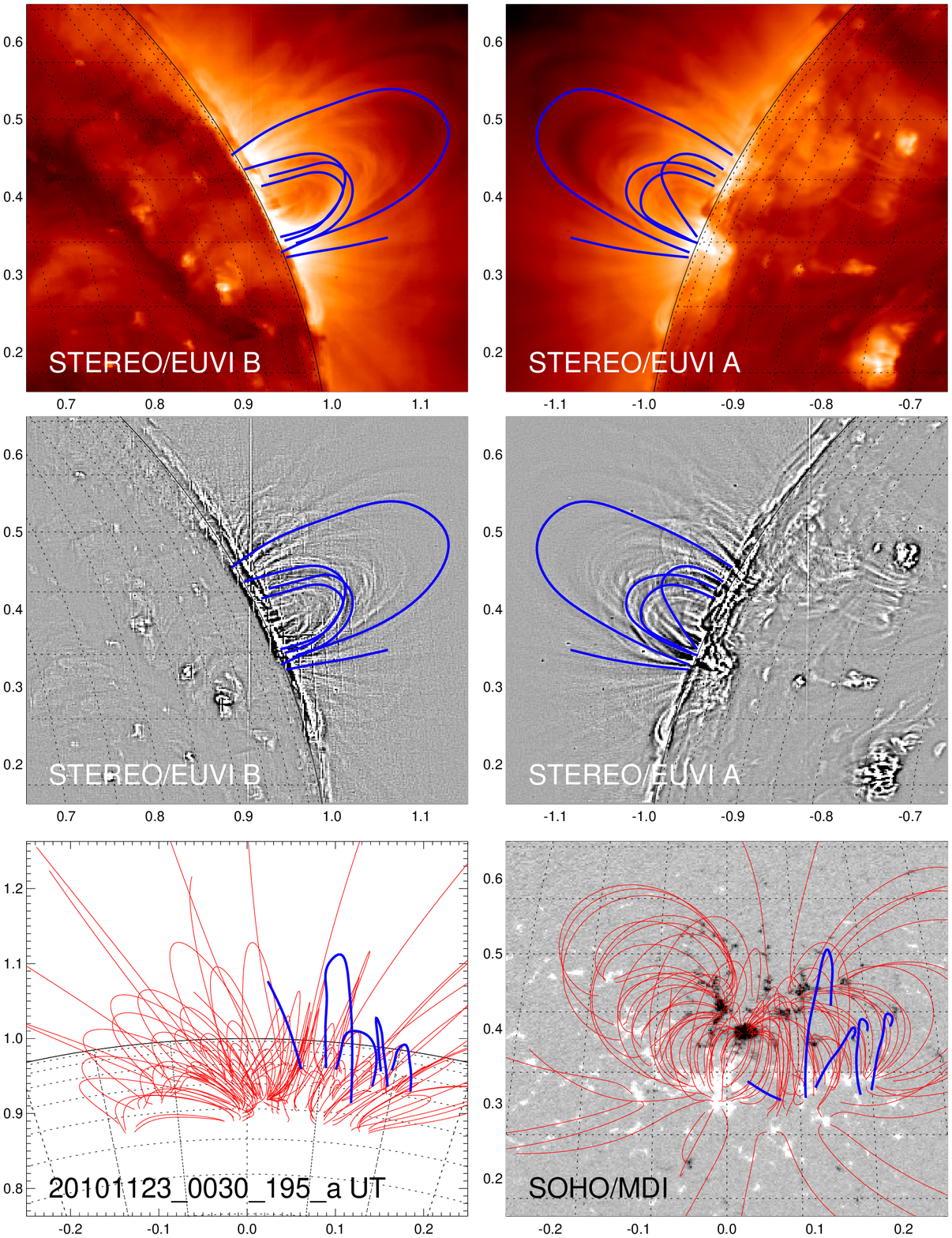}}
\caption{Active region AR 11127, observed on 2010 Nov 23, 00:30 UT 
in the 195 \ang\ wavelength, with a spacecraft separation angle of 
$\alpha_{sep}=169.4^\circ$. A SOHO/MDI magnetogram is shown, 
overlaid with the stereoscopically triangulated loops (blue curves) 
and magnetic field lines computed with a potential field model.
Magnetic field lines have a footpoint threshold of $B>100$ G.}
\end{figure}

The second case is active region NOAA 10978 (Fig.~3), observed on
2007 Dec 11 with a spacecraft separation angle of $\alpha_{sep}=42.7^\circ$.
Note that the views from EUVI/A and B appear already to be significantly 
different with regard to the orientation of the triangulated loops, as
seen from a distinctly different aspect angle. A set of 52 coronal loops
were stereoscopically triangulated in this region 
(Aschwanden and Sandman 2010), 
a mean misalignment angle of $\mu_{PF}=20.8^\circ$ is found for a potential 
field model, and a reduced value of $\mu_{FFF}=15.1^\circ$ is found
for the force-free model (Table 2), while a stereoscopic error of 
$\alpha_{SE}=8.9^\circ$ is estimated (Aschwanden and Sandman 2010). 
Thus, the quality of stereoscopic triangulation
(as well as the degree of non-potentiality) is similar to the first 
active region, although we performed stereoscopy with a 7 times 
larger spacecraft separation angle ($\alpha_{sep}=42.7^\circ$) than before
($\alpha_{sep}=6.1^\circ$). Apparently, stereoscopy is still easy at
such angles, partially helped by the fact that the active region is
located near the central meridian ($\pm 20^\circ$) for both spacecraft, 
which provides an unobstructed view from top down, so that the peripheral 
loops of the active region do not overarch the core of the active region,
where the bright reticulated moss pattern (Berger et al.~1999)
makes it almost impossible 
to discern faint loops in the highpass-filtered images. The top-down
view provides also an optimum aspect angle to disentangle closely-spaced 
loops, which is an important criterion in the stereosopic correspondence 
identification.
  
The third case is active region NOAA 11010 (Fig.~4), observed on
2009 Jan 12 with a near-orthogonal spacecraft separation angle of 
$\alpha_{sep}=89.3^\circ$. Due to the quadrature of the spacecraft,
only a sector of $\pm 45^\circ$ east and west of the central
meridian (viewed from Earth) is jointly visible by both spacecraft. 
This particular active region is seen at a $45^\circ$ angle by both 
STEREO/A and STEREO/B. This symmetric view is the optimum condition
to discern a large number of inclined loop segments and to identify 
the stereoscopic correspondence. We triangulate some
20 loop segments, which appear almost mirrored in the STEREO/A
and B image due to the east-west symmetry of the magnetic dipole.
A mean misalignment angle of $\mu_{PF}=36.2^\circ$ with the
potential field model is found, and a reduced value of
$\mu_{FFF}=16.5^\circ$ with the force-free field model.
An estimate of the statistical (non-parallelity)
stereoscopic error is not possible due to the small number of
triangulated loops. Thus, we conclude that stereoscopy is still
possible in quadrature. Mathematically, the orthogonal projections 
should yield the most accurate 3D coordinates of a curvi-linear 
structure, but in practice, confusion of multiple structures with 
near-aligned projections can cause a disentangling problem in the
stereoscopic correspondence identification at this intermediate angle.

The fourth case is active region NOAA 11032 (Fig.~5), observed on
2009 Nov 21 with a large spacecraft separation angle of 
$\alpha_{sep}=126.9^\circ$. STEREO/A sees the active region near the
east limb from an almost side-on perspective, while STEREO/B sees a
similar mirror image near the west limb, where confusion near 
the limb makes the stereoscopic correspondence identification more 
difficult. We trace some 15 loop segments, but do not succeed
in pinning down a larger number of loops, partially because this active
region is small and does not exhibit numerous bright loops, and partially
because of increasing confusion problems near the limb. We
searched for larger active regions over several months around this time,
but were not successful due to a dearth of solar activity during this time.
We find a misalignment angle of $\mu_{PF}=29.8^\circ$ for the potential
field, and $\mu_{FFF}=17.4^\circ$ for the force-free field model, 
which is still 
comparable with the previous active regions triangulated at smaller 
sterescopic angles. Thus, stereoscopy seems to be still feasible at 
such large stereoscopic angles.

The last case is active region NOAA 11127 (Fig.~6), observed on
2010 Nov 23 with a very large spacecraft separation angle of 
$\alpha_{sep}=169.4^\circ$, only two months before the two STEREO 
spacecraft pass the largest separation point. At this point, the common
field-of-view that is overlapping from STEREO/A and B is only
the central meridian zone seen from Earth (or the opposite meridian
behind the Sun). STEREO/A observes active region NOAA 11127 at
its east limb, while STEREO/B sees it at its west limb, so both
spacecraft see only the vertical structure of the active region
from a side view (see Fig.~6 top). This particular configuration
is very unfavorable for stereoscopy. Although the vertical
structure in altitude can be measured very accurately, the
uncertainty in horizontal direction in longitude is very large
and suffers moreover the sign ambiguity of positive or negative
longitude difference with respect to the limb seen from Earth.
Consequently, we have reliable information on the altitude and
latitude of loops, while the longitude is essentially ill-defined.
In order to reduce the large scatter in the measurement of $z$-coordinates 
along a loop, introduced by the near-infinite amplification
of parallax uncertainties tangentially at the limb, we restrict
the general solution of geometric 3D triangulation to planar loops, 
by applying a linear regression fit of the $z(y)$ coordinates.
The example in Fig.~6 shows that we can trace some (5) loops
in the plane of the sky and have no problem in identifying the
stereoscopic counterparts in both STEREO/A and B images, but the 
stereoscopic triangulation is ill-defined at this singularity 
of the sign change in the parallax effect. The misalignment
between the three loop directions and the potential field is
$\mu_{PF}=39.4^\circ$, and for the force-free field model is
$\mu_{FFF}=21.2^\circ$, which indicates that the 
orientation of the loop planes is less reliably determined.
Stereoscopic triangulation brakes down at this singularity of
separation angles at $\alpha_{sep} \approx 180^\circ$, 
although the stereoscopic correspondence problem is very much 
reduced for the ``mirror images'', similar to the near-identical
images at small separation angles $\alpha_{sep} \gapprox 0^\circ$.

\section{	DISCUSSION				}

We are discussing now the pro's and con's of stereoscopy at small
and large aspect angles, which includes quantitative estimates of
the formal error of stereoscopic triangulation (Section 3.1),
the stereoscopic correspondence and confusion problem (Section 3.2),
and the statistical probability of stereoscopable active regions
during the full duration of the STEREO mission (Section 3.3),
all as a function of the stereoscopic aspect angle (or spacecraft 
separation angle $\alpha_{sep}$ in the case of the STEREO mission). 

\subsection{	Stereoscopic Triangulation Error	}

Stereoscopic triangulation involves a parallax angle around the normal
of the epipolar plane. For the STEREO mission, the epipolar plane intersects
the Sun center and the two spacecraft A and B positions, which are 
separated mostly in east-west direction. No parallax effect occurs when 
the loop axis coincides with the epipolar plane, i.e., when the loop axis points
in east-west direction. Thus, the accuracy of stereoscopic triangulation 
depends most sensitively on this orientation angle $\vartheta$, which we 
define as the angle between the loop direction and the normal of the 
epipolar plane (i.e., approximately 
the y-axis of a solar image in north-south direction). If the position of a
loop centroid can be determined with an accuracy of a half pixel size 
$\Delta_{pix}/2$, 
the dependence of the stereoscopic error on the orientation angle 
$\vartheta$ is then (Aschwanden et al.~2008),
\begin{equation}
	\sigma(\vartheta)={\Delta_{pix} \over 2} \sqrt{1 + \tan^2(\vartheta)} 
	                 ={\Delta_{pix} \over 2} {1 \over \cos(\vartheta)} \ .
\end{equation}
Thus, for a highly inclined loop that 
has an eastern and western footpoint at the same latitude,
the stereoscopic error is minimal near the loop footpoints (pointing in
north-south direction) and at maximum near the loop apex (pointing in
east-west direction).

In addition, the accuracy of stereoscopic triangulation depends also on
the aspect angle (or spacecraft separation angle $\alpha_{sep}$), which
can be quantified with an error trapezoid as shown in Fig.~7. If the
half separation angle $\alpha_{sep}/2$ is defined symmetrically to the
Earth-Sun axis (z-axis), the uncertainty in z-direction is
$\sigma_z=(\Delta_{pix}/2)/\sin(\alpha_{sep}/2)$ and is in the x-direction
(in the epipolar plane)
$\sigma_x=(\Delta_{pix}/2)/\sin([\pi-\alpha_{sep}]/2)$. Including also a
half pixel-size error $\sigma_y=(\Delta_{pix}/2)$ in the y-direction,
we have then a combined error for the 3D position of a triangulated point as,
\begin{equation}
	\sigma(\alpha_{sep})=\sqrt{\sigma_x^2+\sigma_y^2+\sigma_z^2}
	={\Delta_{pix}\over 2}
	\sqrt{1+{1\over \sin^2({\alpha_{sep}/2}) }
	       +{1\over \sin^2{[(\pi-\alpha_{sep})/2]}  }} \ .
\end{equation}
This positional error is symmetric for small $\alpha_{sep}$ and large 
stereoscopic angles $(\pi-\alpha_{sep})$, and has a minimum at an orthogonal
angle of $\alpha_{sep}=\pi/2$. The error is largest in z-direction 
for small spacecraft separation angles, while it is
largest in x-direction for separation angles near $\alpha_{sep} \lapprox
180^\circ$.

\begin{figure}
\centerline{\includegraphics[width=0.5\textwidth]{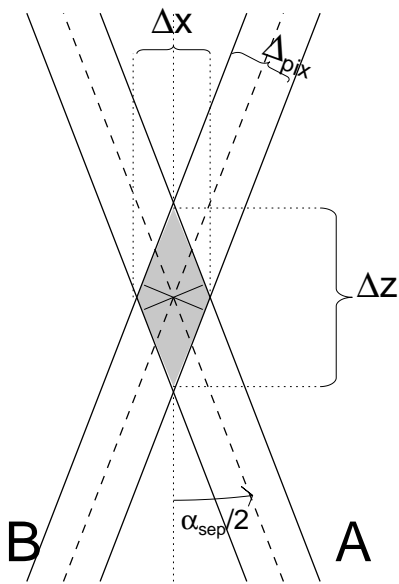}}
\caption{The error trapezoid of stereoscopic triangulation is
shown (grey area), given by the two line-of-sights of the two
observer directions A and B, separated by an angle $\alpha_{sep}$.
The uncertainties $\Delta x$ in x-direction and spacecraft in
z-direction depend on the pixel width $\Delta_{pix}$ and half
aspect angle $\alpha_{sep}/2$.}
\end{figure}

To compare the relative importance of the two discussed sources of errors
we can evaluate the parameters that increase the individual errors by
a factor of two. This is obtained when the orientation angle of a loop 
segment (with respect to the east-west direction) increases from
$\vartheta = 0^\circ$ to $\vartheta \approx 60^\circ$, 
or if the spacecraft separation angle changes from the optimum angle
$\alpha_{sep}=90^\circ$ to $\alpha_{sep}=25^\circ$ 
(or $\alpha_{sep}=155^\circ$, respectively). If stereoscopy at small
angles of $\alpha_{sep}=5^\circ$ is attempted, the positional error
is about ten-fold (corresponding to $\approx 8 \Delta_{pix}$), 
compared with the optimum angle at $\alpha_{sep}=90^\circ$
(corresponding to $\approx 0.8 \Delta_{pix}$). For STEREO/EUVI with a
pixel size of $\Delta_{pix}=1.6\arcsec\approx 1.2$ Mm, this amounts
to an accuracy range of $\approx 1-10$ Mm. 

\subsection{	Stereoscopic Correspondence and Confusion Error }

The previous considerations are valid for isolated loops that can
be unambiguously disentangled in an active region, in both the
STEREO/A and B images. However, this is rarely the case. In crowded
parts of active regions, the correspondence of a particular loop in 
image A with the identical loop in image B can often not properly 
be identified. We call this confusion problem also the {\sl
stereoscopic correspondence problem}, which appears in every 
stereoscopic tie-point triangulation method. In order to quantify
this source of error, we have to consider the area density of loops
and their relative orientation. A top-down view of an active region,
e.g., as seen for small stereoscopic angles by both spacecraft
for an active region near disk center (e.g., Fig.~2), generally
allows a better separation of individual loops, because only the
lowest density scale height is detected (due to hydrostatic
gravitational stratification), and neighbored loop segments do not
obstruct each other due to the foreshortening projection effect
near the footpoints. In contrast, every active region seen near
the limb, shows many loops at different longitudes, but at similar
latitudes, cospatially on top of each other, which represents the
most severe confusion problem. Thus, we can essentially quantify
the degree of confusion by the loop number density per pixel,
which approximately scales with the inverse cosine-function of
the center-to-limb angle due to foreshortening. 
In other words, we can define a quality 
factor $C(\rho)$ for identifying the stereoscopic correspondence 
$C$ of properly disentangled coronal loops, which drops from 
$C(\rho=0)=1$ at disk center to $C(\rho=\pi/2)=0$ at 
the limb, where $\rho$ is the center-to-limb angle measured from 
Sun center, 
\begin{equation}
	C(\rho) = |\cos(\rho)| \qquad 0 \le |\rho| \le \pi/2 \ .
\end{equation}

\begin{figure}
\centerline{\includegraphics[width=0.9\textwidth]{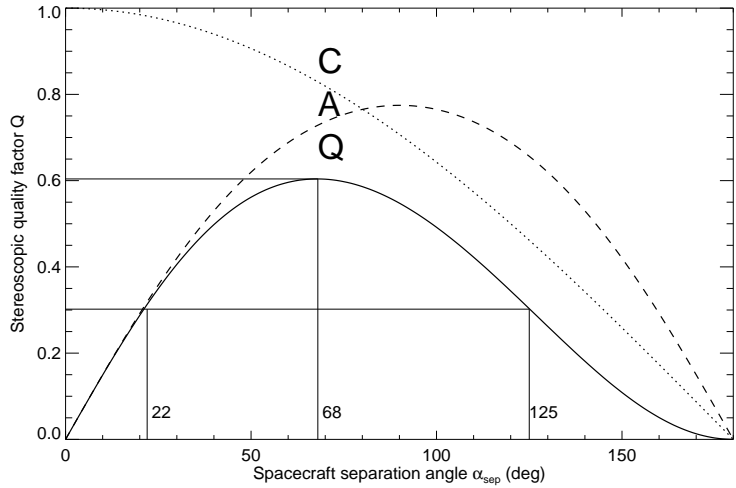}}
\caption{Quality factor $Q$ of stereoscopic triangulation 
as a function of the spacecraft separation angle $\alpha_{sep}$,
which is a function of the accuracy $A$ of triangulated stereoscopic 
positions and the stereoscopic correspondence quality factor $C$.
The best quality (within a factor of 2) occurs in the range of 
$\alpha_{sep}=22^\circ-125^\circ$.}
\end{figure}

The orbits of the STEREO mission reduce the overlapping area on the
solar surface that can be jointly viewed by both spacecraft A and B
linearly with increasing spacecraft separation angle, so that the
center-to-limb distance $\rho$ of an active region located on the central
meridian (viewed from Earth) is related to the spacecraft separation
angle by
\begin{equation}
	\rho = {\alpha_{sep} \over 2} \ ,
\end{equation}
increasing linearly with the separation angle from $\rho=0$ at the
beginning of the mission to $\rho=\pi/2$ at maximum spacecraft
separation angle. The location of an active region at the
central meridian provides the best view for both spacecraft,
because an asymmetric location would move the active region closer
to the limb for one of the spacecraft, and thus would increase the
degree of confusion, as we verified by triangulating a number of
asymmetric cases. Thus, we can express the quality factor of
stereoscopic correspondence $C$ (Eq.~7) by the spacecraft
separation angle (Eq.~8) and obtain the relationship, 
\begin{equation}
	C(\alpha_{sep}) = \cos(\alpha_{sep}/2) 
	\qquad 0 \le \alpha_{sep} \le \pi \ .
\end{equation}

Defining a quality factor $Q$ for stereoscopic triangulation by
combining the stereoscopic correspondence quality $C$ (Eq.~9) 
with the accuracy $A$ of stereoscopic positions (Eq.~6), which 
we may define by the normalized inverse error 
(i.e., $A= \sigma(\alpha_{sep,min})/\sigma(\alpha_{sep})$), we obtain
\begin{equation}
	Q(\alpha_{sep}) = C \times A  
	={ \sqrt{3} \ \cos{(\alpha_{sep}/2)} \over 
	\sqrt{1+{1/\sin^2{(\alpha_{sep}/2)} }
	       +{1/\sin^2{[(\pi-\alpha_{sep})/2}] }}} \ .
\end{equation}
We plot the functional dependence of this stereoscopic quality factor
$Q(\alpha_{sep})$ together with their underlying factors 
$A(\alpha_{sep})$ and $C(\alpha_{sep})$ in Fig.~8 and obtain now an 
asymmetric function of time (or spacecraft separation angle) that 
favors smaller stereoscopic angles. The stereoscopic quality factor 
is most favorable (within a factor of 2) in the range of 
$\alpha_{sep}=22^\circ-125^\circ$, which corresponds to the 
mission phase between August 2007 and November 2009. The same
optimum range will repeat again at the backside of the Sun 
5 years later between August 2012 and November 2014. 

From our analysis of 5 active regions spread over the entire
spacecraft separation angle range we find acceptable results
regarding triangulation accuracy in the range of separation 
angles of $\alpha_{sep}=6^\circ$ and $127^{\circ}$ (based on
acceptable misalignment angles of $\alpha_{PF} \lapprox 35^\circ$,
which coincides with the predicted optimum range of
$\alpha_{sep} \approx 22^\circ-125^\circ$, while stereoscopy
definitely brakes down at $\alpha_{sep} \lapprox 170^\circ$,
as predicted by theory (Fig.~8 and Eq.~10).  

\begin{figure}
\centerline{\includegraphics[width=0.9\textwidth]{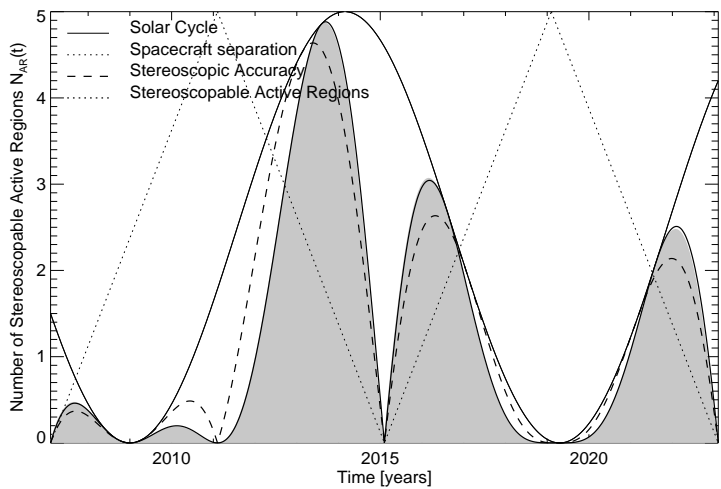}}
\caption{The statistically expected variation of the number of 
active regions $N(t)$ during the solar cycle (thin solid curve),
the spacecraft separation angle $\alpha_{sep}(t)$ (dotted curve), 
the stereoscopic accuracy $A(t)$ (dashed curve), and the expected 
number of stereoscopable active regions $N_{AR}(t)$ (curve with
grey area) as a function of time during a full 16-year mission cycle 
of the STEREO mission.}
\end{figure}

\subsection{	The Statistical Probability of Stereoscopy	}

There are different factors that affect the quality or feasibility
of solar stereoscopy, such as (i) the availability of large active regions
(which varies statistically as a function of solar cycle), (ii) the
simultaneous viewing by both spacecraft STEREO/A and B (which depends
on the spacecraft separation angle), (iii) the geometric foreshortening
that affects the stereoscopic correspondence problem 
(which depends on the center-to-limb distance for each spacecraft view), 
and (iv) the time of the central meridian passage of the active region 
for a viewer from Earth (which determines the symmetry of views for both
spacecraft, where minimum confusion occurs in the stereoscopic 
correspondence identification). All but the first factor depend on
the spacecraft separation angle $\alpha_{sep}(t)$, which is a specific
function of time for the STEREO mission (with a complete cycle of 16
years). In order to assess the science return of the STEREO mission
or future missions with stereoscopic capabilities, it is instructive
to quantify the statistical probability of acceptable stereoscopic
results as a function of spacecraft separation angle or time.

We already quantified the quality of stereoscopy $Q(\alpha_{sep})$
as a function of the stereoscopic angle in Eq.~(10). Let us define
the number probability $N(t)$ of existing active regions at a given 
time $t$ with a squared sinusoidal modulation during the solar cycle,
\begin{equation}
	N(t) = N_0\ \sin{\left( {\pi (t-t_0) \over T_{cycle}}
	\right) } \ ,
\end{equation}
where $N_0 \approx 10$ is the maximum number of active regions
existing on the total solar surface during the maximum of the solar cycle,
$t_0$ is the time of the solar minimum (e.g., $t_{min} \approx 2009$),
and $T_{cycle} \approx 10.3$ yrs the current average solar cycle length.

The second effect is the overlapping area on the solar surface that
is simultaneously seen by both spacecraft STEREO/A and B, which 
decreases linearly with the spacecraft separation angle from 50\%
at $\alpha_{sep}(t=t_1)$ (with $t_1=2007.1$ at the start of spacecraft
separation) to 0\% at $\alpha_{sep}(t=t_2)$ (with $t_2=2011.1$ at
maximum separation), and then increases linearly again for the next
quarter phase of a mission cycle. If we fold the variation $N(t)$ 
of the solar cycle (Eq.~11) with the triangular stereoscopic overlap
area variation $A(t)$ together, we obtain a statistical probability
for the number of stereoscopically triangulable active regions.
However, the number of accurate stereoscopical triangulations scales
with the quality factor $Q(\alpha_{sep})$ (Eq.~10), where the
spacecraft separation angle $\alpha_{sep}(t)$ is a piece-wise linear 
(triangular) function of time $t$ according to the spacecraft orbit.
Essentially we are assuming that the probability of successfull 
stereoscopic triangulations at a given time scales with the quality 
factor or feasibility of accurate stereoscopy at this time.
So, we obtain a combined probability of stereoscopically triangulable
active regions of,
\begin{equation}
	N_{AR}(t) = N(t) \times Q[\alpha_{sep}(t)] \ .
\end{equation}
In Fig.~9 we show this combined statistical probability of 
feasible stereoscopy in terms of the expected number of active
regions for a full mission cycle of 16 years, from 2006 to 2022. 
It shows that the best periods for solar stereoscopy are during
2012-2014, 2016-2017, and 2021-2023.

\section{	CONCLUSIONS			}

After the STEREO mission reached for the first time a full $360^\circ$
view of the Sun this year (2007 Feb 6), the two STEREO A and B spacecraft
covered also for the first time the complete range of stereoscopic
viewing angles from $\alpha_{sep} \gapprox 0^\circ$ to 
$\alpha_{sep} \lapprox 180^\circ$. We explored the feasibility of
stereoscopic triangulation for coronal loops in the entire angular
range by selecting 5 datasets with viewing angles at 
$\alpha_{sep} \approx 6^\circ, 43^\circ, 89^\circ, 127^\circ$ and 
$169^\circ$. Because previous efforts for solar stereoscopy covered
only a range of small stereoscopic angles ($\alpha_{sep} \lapprox 45^\circ$),
we had to generalize the stereoscopic triangulation code for large angles
up to $\alpha_{sep} \le 180^\circ$. We find that stereoscopy of coronal
loops is feasible with good accuracy for cases in the range $\alpha_{sep}
\approx 6^\circ - 127^\circ$, a range that is also theoretically 
predicted by taking into account the triangulation errors due to 
finite spatial resolution and confusion in the stereoscopic correspondence 
identification in image pairs, which is hampered by projection effects and
foreshortening for viewing angles near the limb. Accurate stereoscopy
(within a factor of 2 of the best possible accuracy) is predicted for
a spacecraft separation angle range of 
$\alpha_{sep} \approx 22^\circ - 125^\circ$.  
Based on this model we predict that the best periods for stereoscopic
3D reconstruction during a full 16-year STREREO mission cycle occur
during 2012-2014, 2016-2017, and 2021-2023, taking the variation in the
number of active regions during the solar cycle into account also.

Why is the accuracy of stereoscopic 3D reconstruction so important?
Solar stereoscopy has the potential to quantify the coronal magnetic
field independently of conventional 2D magnetogram and 3D vector
magnetograph extrapolation methods, and thus serves as an important
arbiter in testing theoretical models of magnetic potential fields,
linear force-free field models (LFFF), and nonlinear force-free
field models (NLFFF). A benchmark test of a dozen of NLFFF codes
has been compared with stereoscopic 3D reconstruction of coronal
loops and a mismatch in the 3D misalignment angle of $\mu \approx
24^\circ-44^\circ$ has been identified (DeRosa et al.~2009), which
is attributed partially to the non-force-freeness of the photospheric
magnetic field, and partially to insufficient constraints of the
boundary conditions of the extrapolation codes. Empirical estimates
of the error of stereoscopic triangulation based on the non-parallelity
of loops in close proximity has yielded uncertainties of 
$\mu_{SE} \approx 7^\circ-12^\circ$. Thus the residual difference in
the misalignment is attributed to either the non-potentiality of the
magnetic field (in the case of potential field models), or to the
non-force-freeness of the photospheric field (for NLFFF models).
We calculated also magnetic potential fields here for all
stereoscopically triangulated active regions and found mean
misalignment angles of $\mu_{PF}\approx 21^\circ-39^\circ$,
which improved to $\mu_{FFF}\approx 15^\circ-21^\circ$ for a
nonlinear force-free model, which testifies
the reliability of stereoscopic reconstruction for the first time
over a large angular range. The only case where stereoscopy clearly fails
is found for an extremely large separation angle of  
($\alpha_{sep}\approx 170^\circ$), which is also reflected in the largest
deviation of misalignment angles found ($\mu_{NP} \approx 39^\circ$,
$\mu_{FFF} \approx 21^\circ$).
Based on these positive results of stereoscopic accuracy over an 
extended angular range from small to large spacecraft separation angles
we anticipate that 3D reconstruction of coronal loops by stereoscopic
triangulation will continue to play an important role in testing
theoretical magnetic field models for the future phases of the STEREO 
mission, especially since stereoscopy of a single image pair does not 
require a high cadence and telemetry rate at large distances behind the Sun.

\acknowledgements
This work is supported by the NASA STEREO mission 
under NRL contract N00173-02-C-2035.
The STEREO/ SECCHI data used here are produced by an international consortium of
the Naval Research Laboratory (USA), Lockheed Martin Solar and Astrophysics Lab
(USA), NASA Goddard Space Flight Center (USA), Rutherford Appleton Laboratory
(UK), University of Birmingham (UK), Max-Planck-Institut f\"ur
Sonnensystemforschung (Germany), Centre Spatiale de Li\`ege (Belgium), Institut
d'Optique Th\'eorique et Applique (France), Institute d'Astrophysique Spatiale
(France).
The USA institutions were funded by NASA; the UK institutions by
the Science \& Technology Facility Council (which used to be the Particle
Physics and Astronomy Research Council, PPARC); the German institutions by
Deutsches Zentrum f\"ur Luft- und Raumfahrt e.V. (DLR); the Belgian institutions
by Belgian Science Policy Office; the French institutions by Centre National
d'Etudes Spatiales (CNES), and the Centre National de la Recherche Scientifique
(CNRS). The NRL effort was also supported by the USAF Space Test Program and
the Office of Naval Research.

\section*{References} 

\def\ref#1{\par\noindent\hangindent1cm {#1}}

\small
\ref{Aschwanden, M.J., W\"ulser, J.P., Nitta, N., Lemen, J. 2008, 
	\apj {\bf 679}, 827.}
\ref{Aschwanden, M.J. 2009, \ssr {\bf 149}, 31.}
\ref{Aschwanden, M.J. and Sandman, A.W. 2010, \aj {\bf 140}, 723.}
\ref{Aschwanden, M.J. 2012, Sol.Phys. (subm.), 
        {\sl A nonlinear force-free magnetic field approximation suitable for
        fast forward-fitting to coronal loops. I. Theory},
        \url{http:www.lmsal.com/~aschwand/eprints/2012_fff1.pdf}}
\ref{Aschwanden, M.J. and Malanushenko,A. 2012, Sol.Phys. (subm),
        {\sl A nonlinear force-free magnetic field approximation suitable for
        fast forward-fitting to coronal loops. II. Numeric Code and Tests},
        \url{http://www.lmsal.com/~aschwand/eprints/2012_fff2.pdf}}
\ref{Aschwanden, M.J., Wuelser,J.-P., Nitta, N.V., Lemen, J.R., Schrijver,
        C.J., DeRosa, M., and Malanushenko, A. 2012, ApJ (subm),
        {\sl First 3D Reconstructions of Coronal Loops with the STEREO A and B
        Spacecraft: IV. Magnetic Field Modeling with Twisted Force-Free
        Fields},
        \url{http://www.lmsal.com/~aschwand/eprints/2012_stereo4.pdf},
        \url{http://www.lmsal.com/~aschwand/movies/STEREO_fff_movies}}
\ref{Bemporad, A. 2009, \apj {\bf 701}, 298.}
\ref{Berger, T.E., DePontieu, B., Fletcher, L., Schrijver, C.J., 
	Tarbell, T.D., and Title, A.M. 1999, \sp {\bf 190}, 409.}
\ref{DeRosa, M.L., Schrijver, C.J., Barnes, G., Leka, K.D., Lites, B.W., 
	Aschwanden, M.J., Amari, T., Canou, A., McTiernan, J.M., Regnier, S., 
	Thalmann, J., Valori, G., Wheatland, M.S., Wiegelmann, T., 
	Cheung, M.C.M., Conlon, P.A., Fuhrmann, M., Inhester, B., 
	and Tadesse, T. 2009, \apj {\bf 696}, 1780.}
\ref{Feng, L., Inhester, B., Solanki, S., Wiegelmann, T., Podlipnik, B., 
	Howard, R.A., and W\"ulser, J.P. 2007, \apj {\bf 671}, L205.}
\ref{Feng, L., Inhester, B., Solanki, S.K., Wilhelm, K., Wiegelmann, T., 
	Podlipnik, B., Howard, R.A., Plunkett, S.P., W\"ulser, J.P., and
	Gan, W.Q. 2009, \apj {\bf 700}, 292.}
\ref{Howard, R.A., Howard,R.A., Moses, J.D., Vourlidas, A., Newmark, J.S., 
	Socker, D.G., Plunkett, S.P., Korendyke, C.M., Cook, J.W., Hurley, A., 
	Davila, J.M. and 36 co-authors, 2008, \ssr {\bf 136}, 67.}
\ref{Inhester, B. 2006, ArXiv e-print: astro-ph/0612649.}
\ref{Kaiser, M.L., Kucera, T.A., Davila, J.M., St. Cyr, O.C., 
	Guhathakurta, M., and Christian, E. 2008, \ssr {\bf 136}, 5.}
\ref{Liewer, P.C., DeJong, E.M., Hall, J.R., Howard, R.A., Thompson, W.T., 
	Culhane, J.L., Bone, L., van Driel-Gesztelyi, L. 2009, 
 	\sp {\bf 256}, 57.}
\ref{Sandman, A., Aschwanden, M.J., DeRosa, M., W\"ulser, J.P., and 
	Alexander, D. 2009, \sp {\bf 259}, 1.}
\ref{Sandman, A.W. and Aschwanden, M.J. 2011, \sp {\bf 270}, 503.} 
\ref{Thompson, W.T. 2006, \aap {\bf 449}, 791.}
\ref{Thompson, W.T. 2011, \jastp {\bf 73}, 1138.}
\ref{Thompson, W.T., Davila, J.M., St. Cyr, O.C., and Reginald, N.L.
 	2011, \sp {\bf 272}, 215.}
\ref{W\"ulser, J.P., Lemen, J.R., Tarbell, T.D., Wolfson, C.J., Cannon, J.C., 
	Carpenter, B.A., Duncan, D.W., Gradwohl, G.S., Meyer, S.B., 
	Moore, A.S., and 24 co-authors, 2004, SPIE {\bf 5171}, 111.}
\end{article}
\end{document}